\documentclass[aps,pra,showpacs,floatfix,superscriptaddress,twocolumn]{revtex4}
\usepackage[dvips]{graphicx}

\usepackage{longtable}
\usepackage{dcolumn}
\usepackage[dvips]{graphicx}
\usepackage{bm}
\usepackage{bbm}

\usepackage{times}
\usepackage{nicefrac}
\usepackage{amsmath}
\usepackage{amsfonts}
\usepackage{amssymb}
\usepackage{amsthm}

\newcolumntype{.}{D{x}{}{-1}}

\newcommand{\balpha}{\bm{\alpha}}

\newcommand{\bnabla}{\bm{\nabla}}

\newcommand{\vare}{\varepsilon}
\newcommand{\bfr}{{\bm {r}}}
\newcommand{\bfx}{{\bm {x}}}

\newcommand{\bfp}{{\bm {p}}}

\newcommand{\bfk}{{\bm {k}}}

\newcommand{\lbr}{\langle}
\newcommand{\rbr}{\rangle}

\newcommand{\pr}{^{\prime}}

\begin{document}

\title{Relativistic calculations of double $\bm K\!$-shell photoionization cross sections for neutral
       medium-$\bm Z$ atoms}

\author{V. A. Yerokhin}

\affiliation{Center for Advanced Studies, St.~Petersburg State Polytechnical University,
 195251 St.~Petersburg, Russia}

\affiliation{Helmholtz-Institut Jena, D-07743 Jena, Germany}

\author{A. Surzhykov}

\affiliation{Helmholtz-Institut Jena, D-07743 Jena, Germany}

\author{S. Fritzsche}

\affiliation{Helmholtz-Institut Jena, D-07743 Jena, Germany}
\affiliation{{Theoretisch-Physikalisches Institut, Friedrich-Schiller-Universit\"at Jena, D-07743
Jena, Germany}}

\begin{abstract}

Fully relativistic calculations are presented for the double $K$-shell photoionization cross
section for several neutral medium-$Z$ atoms, from magnesium ($Z = 10$) up to silver ($Z = 47$).
The calculations take into account all multipoles of the absorbed photon as well as the
retardation of the electron-electron interaction. The approach is based on the partial-wave
representation of the Dirac continuum states and uses the Green-function technique to represent
the full Dirac spectrum of intermediate states. The method is strictly gauge invariant, which is
used as an independent cross check of the computational procedure. The calculated ratios of the
double-to-single $K$-shell ionization cross sections are compared with the experimental data and
with previous computations.

\end{abstract}


\maketitle

\section{Introduction}

Double photoionization is a fundamental atomic process in which a \textit{single} photon, being
absorbed by an atom, simultaneously kicks out {\em two} electrons. The characteristic feature of
this process is that it can proceed only through the electron-electron interaction. Indeed, the
incoming photon can (formally) interact only with one electron, so for the second electron to be
kicked out, the required energy should be transferred from one electron to another through the
electron-electron interaction. This feature makes the double photoionization very sensitive to the
details of the electron-electron interaction.

A typical system for studying double photoionization has long been the helium atom, for which
numerous experimental and theoretical investigations have been performed \cite{briggs:00}. The most
widely studied quantity was the ratio of the double-to-single photoionization cross sections
$\sigma^{++}/\sigma^+$ as a function of the energy of the incoming photon $\omega$. Today, the
helium double photoionization is often considered as well understood, and the results of several
independent computations \cite{tang:95,kehifets:96,meyer:97,qiu:98} agree reasonably well with the
experimental data \cite{levin:96,doerner:96,levin:93,samson:98} in the whole region of the photon
energies.

Unlike the helium case, an adequate description of double photoionization of both $K$-shell
electrons in neutral medium- and high-$Z$ atoms is much more challenging for theory. This is
because of the enhanced relativistic effects (that scale as $Z^2$) as well as the more complex
electronic structure of many-electron atoms. The difficulties of the theory in this case were
demonstrated by experiments of Kantler {\em et al.} on molybdenum \cite{kantler:99} and later also
silver \cite{kantler:06}, which reported large disrepancies with nonrelativistic calculations and
claimed ``the need for theoretical treatments to properly deal with such systems". More recently,
Hoszkowska {\em et al.} \cite{hoszowska:09,hoszowska:10} presented detailed experimental
investigations of the double $K$-shell photoionization for eight medium-$Z$ atoms in the range $12
\le Z \le 23$. In the absence of suitable {\em ab initio} calculations, the experimental data of
Ref.~\cite{hoszowska:09,hoszowska:10} were interpreted only in terms of various semi-empirical
models.

An attempt for a systematic {\em ab inito} calculation of the double $K$-shell photoionization in
neutral atoms was reported by Kheifets {\em et al.} \cite{kheifets:09}. In that work,
nonrelativistic close-coupling calculations were performed with three gauges of the electromagnetic
operator: length, velocity, and acceleration. Convergence (or the lack of it) between the
calculations in the different gauges is commonly used as a test of the accuracy of the treatment of
the electron correlation. The calculations of Ref.~\cite{kheifets:09} showed that deviations of
double photoionization results obtained in different gauges are large and becoming even larger as
the nuclear charge and/or the photon energy are increased. The authors \cite{kheifets:09} concluded
that ``significant difficulties" arise and that ``none of the available ground-state wave functions
satisfied the strict gauge convergence test".

In a previous study of two of us \cite{yerokhin:11:dpi}, a calculation of the double
photoionization was performed within a fully relativistic framework, for which the problem of gauge
dependence does not arise as the formalism is gauge invariant from the very beginning. Large
relativistic effects were demonstrated in that work, but results were reported only for the He-like
ions. In the present investigation, we extend our previous approach to many-electron systems and
present fully relativistic calculation of the double $K$-shell photoionization cross section in
neutral atoms.

The computational approach is based on the relativistic QED perturbation theory. It takes into
account all multipoles of the absorbed photon as well as the retardation (the frequency dependence)
of the electron-electron interaction. The electron-electron interaction is accounted for rigorously
to the leading order of perturbation expansion. The higher-order electron-electron interaction (in
particular, the interaction with the spectator electrons) is taken into account approximately, by
means of a suitable screening potential in the Dirac equation.

The remaining paper is organized as follows. In Sec.~\ref{sec:theory} we briefly describe our
theoretical approach. Sec.~\ref{sec:numerics} presents details of the calculation. Numerical
results are presented and discussed in Sec.~\ref{sec:res}. The relativistic units ($\hbar=c=m=1$)
are used throughout this paper.

\section{Theory}
\label{sec:theory}

We consider the process, in which an incoming photon with energy $\omega$ and helicity $\lambda$
collides with a neutral atom and kicks out {\em two} electrons from the $K$-shell into the
continuum. The final-state electrons have the energies $\vare_1$ and $\vare_2$ and the momentum
$\bfp_1$ and $\bfp_2$, respectively. Such a process can occur when the photon energy $\omega$ is
equal to or greater than the threshold energy $\omega_{\rm cr}$, which is the (double) ionization
energy of the $K$ shell.

The  energy-differential cross section of the double $K$-shell photoionization is given by
\cite{yerokhin:11:dpi}
\begin{align}  \label{e124}
\frac{d\sigma^{++}}{d\vare_1} =
 \frac{4\pi^2\alpha}{\omega}\,\sum_{\kappa_1\kappa_2\mu_1\mu_2}
    \left| \tau^{++}_\lambda(\vare_1\kappa_1\mu_1,\vare_2\kappa_2\mu_2;\omega)\right|^2\,,
\end{align}
where $\tau_{\lambda}^{++}$ is the amplitude of the process. The summation in the above formula
runs over the partial waves of the continuum wave functions of the ejected (final-state) electrons,
i.e., the relativistic angular quantum numbers $\kappa_{1,2}$ and the projections of the total
angular momentum $\mu_{1,2}$. The remaining $N-2$ (spectator) electrons do not change their state
during the ionization process.

Since the two outgoing electrons share the excess energy of the photoionizataion process, the {\em
total cross section} is obtained as the integral of the single differential cross section over a
half of the energy sharing interval
\begin{align} \label{eq:total}
\sigma^{++} &\ = \int_{m}^{m+(\omega-\omega_{\rm cr})/2}d\vare_1\,
\frac{d\sigma^{++}}{d\vare_1}\,,
\end{align}
where $\omega_{\rm cr}$ denotes the threshold of the double photoionization process. The other half
of the energy interval corresponds to interchanging the first and the second electron and is
accounted for by the antisymmetrized electron wave function.

\begin{figure}[tb]
  \centerline{\includegraphics[width=\columnwidth]{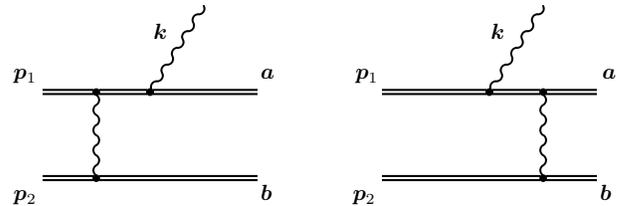}}
\caption{Feynman diagrams that represent the double photoionization process in the
  leading order   of perturbation theory. $a$ and $b$ denote the bound electron states,
  $\bm{p}_1$ and   $\bm{p}_2$ are the continuum electron states, $\bm{k}$ refers to
  the incoming photon. Double lines denote electrons that are propagating
  in a central potential (nuclear Coulomb field plus some screening potential).
  Proper antisymmetrization of the initial and
  the final two-electron wave functions is assumed.
  \label{fig:dpi} }
\end{figure}

To the leading order of QED perturbation theory, the amplitude of the double photoionization
process is represented by the two Feynman diagrams as displayed in Fig.~\ref{fig:dpi}. The general
expression for the amplitude was derived in Ref.~\cite{yerokhin:11:dpi} by using the two-time
Green's function method \cite{shabaev:02:rep},
\begin{widetext}
\begin{align}  \label{e2}
 \tau^{++}_{\lambda}
  &\ =
 N\,\sum_{\mu_a\mu_b} C^{J_0M_0}_{j_a \mu_a\,j_b\mu_b}\,
\sum_{PQ}(-1)^{P+Q}\,
\nonumber \\
&\  \times
\sum_n \biggl\{
\frac{\lbr P\vare_1\,P\vare_2|I(\Delta_{P\vare_2\,Qb})|n\,Qb\rbr\, \lbr n|
  R_{\lambda}|Qa\rbr} {\vare_{Qa}+\omega-\vare_n(1-i0)}
 +
\frac{\lbr P\vare_1| R_{\lambda}|n\rbr\,
 \lbr n\,P\vare_2|I(\Delta_{P\vare_2\,Qb})|Qa\,Qb\rbr}
     {\vare_{P\vare_1}-\omega-\vare_n(1-i0)}
 \biggl\}\,.
\end{align}
\end{widetext}
The operators in the matrix elements in the above formula are the frequency-dependent
electron-electron interaction operator $I(\Delta)$ and the operator of the photon absorption
$R_{\lambda}$. The states $|a\rbr\equiv |\kappa_a\mu_a\rbr$ and $|b\rbr\equiv |\kappa_b\mu_b\rbr$
describe the initial bound electron states, whereas $|\vare_1\rbr\equiv |\vare_1\kappa_1\mu_1\rbr$
and $|\vare_2\rbr\equiv |\vare_2\kappa_2\mu_2\rbr$ describe the final continuum electron states
with a definite total angular momentum.

The first term in the curly brackets of Eq.~(\ref{e2}) corresponds to the diagram with the
electron-electron interaction attached to the final-state electron wave function (the left graph in
Fig.~\ref{fig:dpi}) and the second one, to the diagram with the electron-electron interaction
attached to the initial-state electron wave function (the right graph in Fig.~\ref{fig:dpi}). The
summation over $P$ and $Q$  in Eq.~(\ref{e2}) corresponds to the permutation of the initial and
final electrons, $P\vare_1P\vare_2 = (\vare_1\vare_2)$ or $(\vare_2\vare_1)$, $QaQb = (ab)$ or
$(ba)$, and $(-1)^P$ and $(-1)^Q$ are the permutations sign. The summation over $n$  in
Eq.~(\ref{e2}) runs over the complete Dirac spectrum of intermediate states \cite{shabaev:02:rep},
$\Delta_{\vare_i b} \equiv \vare_i-\vare_b$, $N = 1/\sqrt{2}$ for the equivalent initial-state
electrons and $N=1$ otherwise, $j_{a,b}$ and $\mu_{a,b}$ are the total angular momentum and its
projection of the initial-state electrons, and $J_0$ and $M_0$ are the total angular momentum of
the initial two-electron state and its projection. In the case of $K$ shell, $N = 1/\sqrt{2}$, $J_0
= M_0 = 0$, $j_a = j_b = 1/2$.

The general relativistic expression for the photon absorption operator $R_{\lambda}$ is given by
\begin{align} \label{1aa}
R_{\lambda} = \balpha\cdot\hat{\bm{u}}_{\lambda}\,
e^{i\bfk\cdot\bfr}\, + G\,\bigl(\balpha\cdot\hat{\bfk}-1 \bigr) \,e^{i\bfk\cdot\bfr}\,,
\end{align}
where $\balpha$ is a three-component vector of the Dirac matrices, $\hat{\bm{u}}_{\lambda}$ is the
polarization vector of the absorbed photon, $\bfk$ is the photon momentum, $\hat{\bfk} =
\bfk/|\bfk|$, and $G$ is the gauge parameter.

The relativistic frequency-dependent electron-electron interaction operator in the Feynman gauge is
given by
\begin{align} \label{fey}
I^{\rm Feyn}(\omega) = \alpha\,
(1-\balpha_1\cdot\balpha_2)\,\frac{e^{i|\omega|x_{12}}}{x_{12}}\,,
\end{align}
where $x_{12} = |\bfx_1-\bfx_2|$. In the Coulomb gauge, the electron-electron interaction acquires
an additional term,
\begin{align} \label{cou}
I^{\rm Coul}(\omega)&\  = I^{\rm Feyn}(\omega)
\nonumber \\ &
+ \alpha\,
 \left[ 1-
   \frac{(\balpha_1\cdot\bnabla_1)(\balpha_2\cdot\bnabla_2)}{\omega^2}
\right]\,
\frac{1-e^{i|\omega|x_{12}}}{x_{12}}\,.
\end{align}

In our approach, all one-electron states in Eq.~(\ref{e2}) $|\vare_{1,2}\rbr$, $|a\rbr$, $|b\rbr$,
and $|n\rbr$ are assumed to be eigenstates of the same one-particle Dirac Hamiltonian
\begin{equation}\label{eq2}
    h_D = \balpha\cdot\bfp+ (\beta-1)\,m+ V_{\rm nuc}(r) + V_{\rm scr}(r)\,,
\end{equation}
where $\beta$ is the Dirac $\beta$ matrix, $\bfp$ is the momentum operator, $V_{\rm nuc}$ is the
binding potential of the nucleus, and $V_{\rm scr}$ is the screening potential induced by the
presence of other electrons. So, our approach includes the electron-electron interaction to the
first order of the QED perturbation theory exactly, whereas the higher-order electron-electron
interactions are accounted for approximately, through the screening potential in the Dirac
equation. By varying the definition of the screening potential, we can estimate the residual
electron-correlation effects that are omitted in the present treatment.

In the present work, we construct the screening potential by first solving the Dirac-Fock equation
for the neutral atom and then generating the potential as it arises from the charge density of the
Dirac-Fock orbitals weighted by the occupation numbers of the orbitals. In particular, we make use
of two variants of the potential which will be termed as the core-Hartree (CH) potentials $V_{{\rm
CH},1}(r)$ and $V_{{\rm CH},2}(r)$ and which are defined by
\begin{align}\label{eqq4}
    V_{{\rm CH},K}(r)&\  = \alpha \int_0^{\infty} dr\pr \frac1{\max(r,r\pr)}\,
    \nonumber \\ & \times
        \sum_{n} \bigl(q_n- K \,\delta_{na}\bigr)\, \left[ G_n^2(r)+F_n^2(r)\right]\,,
\end{align}
where $K=1$ or $2$, $n$ numerates the one-electron orbitals, $q_n$ is the occupation number of the
orbital, $a$ is the initial $1s$ electron state, and $G_n$ and $F_n$ are the upper and the lower
radial components of the Dirac-Fock orbitals.

As can be seen from the definition (\ref{eqq4}), $V_{{\rm CH},2}(r)$ represents the potential
generated solely by the charge density of the spectator electrons that do not change their state
during the process, whereas $V_{{\rm CH},1}(r)$ includes in addition the interaction with the
second $1s$ electron in the $K$ shell.

It is important that all initial and intermediate states in our approach are the exact eigenstates
of the same one-particle Dirac Hamiltonian $h_D$ with the potential $V_{\rm nuc}(r) + V_{\rm
scr}(r)$. Because of this and the fact that the screening potential $V_{\rm scr}$ constructed by
Eq.~(\ref{eqq4}) is a {\em local} potential, the amplitude (\ref{e2}) is {\em gauge invariant}. We
note that if we had used a nonlocal (e.g., Dirac-Fock) screening potential in $h_D$, it would have
broken the gauge invariance \cite{hata:84}.

It can be proved that the amplitude (\ref{e2}) is separately gauge invariant with respect to the
gauge of the absorbed photon as well as the gauge of the electron-electron interaction. This gauge
invariance was used in order to check our numerical procedure. In addition, the gauge invariance
provides us a cross-check of the computation of the two Feynman diagrams against each other.
Indeed, while the contributions of the two diagrams in Fig.~\ref{fig:dpi} are different for
different gauges of the emitted photon, their sum should be (and was checked numerically to be) the
same.

%
%

\section{Numerical issues}
\label{sec:numerics}

A general (numerical) scheme for the computation of double photoionization cross sections was
developed in our previous investigation \cite{yerokhin:11:dpi}. Similarly to that, the summation
over the complete spectrum of the Dirac equation is performed in the present work by using the
Dirac-Green function. The major new feature of the present computation is that the initial and
final states as well as the Dirac Green function need to be calculated not only for the
point-nucleus Coulomb potential (as in Ref.~\cite{yerokhin:11:dpi}), but for some general
(Coulomb+screening) potential. For the wave functions, such a generalization is quite
straightforward, and these functions can be readily obtained, e.g., with the help of the Fortran
package developed by Salvat et al.~\cite{salvat:95:cpc}. In contrast, an efficient computation of
the Dirac Green function for the general potential is more difficult.

For a given value of the relativistic angular quantum number $\kappa$, the radial part of the Dirac
Green function is represented in terms of the two-component solutions of the radial Dirac equation
that are regular at the origin $\left(\phi_{\kappa}^{0}\right)$ and at infinity
$\left(\phi_{\kappa}^{\infty}\right)$,
\begin{align}\label{gr01}
 G_{\kappa}(E,r_1,r_2) = &\,
 -\phi_{\kappa}^{\infty}(E,r_1)\,\phi_{\kappa}^{0^T}(E,r_2)\,\theta(r_1-r_2)
\nonumber \\ &
 -\phi_{\kappa}^{0}(E,r_1)\,\phi_{\kappa}^{{\infty}^T}(E,r_2)\,\theta(r_2-r_1)\,,
\end{align}
where $E$ denotes the energy argument of the Green function, $r_1$ and $r_2$ are the radial
arguments, and $\theta$ is the Heaviside step function.

An efficient numerical scheme for computing the regular and irregular solutions of the Dirac
equation for an arbitrary Coulomb $+$ short-range potential was developed by one of us in
Ref.~\cite{yerokhin:11:fns}. We note that practical computations for arbitrary potential become
much more time consuming as compared to computations with the point-nucleus Coulomb potential. The
reason is that the point-nucleus Dirac Coulomb Green function can be calculated directly at any
radial point, whereas for arbitrary potential, a two-step procedure is required: first one needs to
solve the radial Dirac equation on the grid and only then one can obtain the regular and irregular
Dirac solutions by interpolation.

A serious numerical problem arises in the evaluation of the radial integrals for the left graph in
Fig.~\ref{fig:dpi} (where the electron correlation modifies the final-state wave function). In this
case, the continuum-state Dirac wave function has to be integrated together with the Dirac Green
function with the energy $E>m$. All functions under the integral are strongly oscillating and
decrease only (very) slowly for large radial arguments. It is therefore practically impossible to
evaluate such integral to high precision by just applying a straightforward numerical integration.
In our approach, we use instead the method of the complex-plane rotation of the integration
contour, described in detail in Ref.~\cite{yerokhin:10:bs}.

After the radial integrals are successfully evaluated, the next problem is the summation of the
partial-wave expansions. When all angular momentum selection rules are taken into account, two (out
of five) partial-wave summations remain infinite. One can choose the parameters for these two
infinite summations differently. Our choice was the relativitic angular momentum parameters
$\kappa_1$ and $\kappa_2$ of the two final-state electrons in Eq.~(\ref{e124}). In the present
work, the number of partial waves included was $(|\kappa_1|,|\kappa_2|) = (10,10)$, with the tail
of the expansion estimated by extrapolation. This is slightly less than in our previous
investigation \cite{yerokhin:11:dpi}, since the maximal photon energy is smaller in the present
work.

A number of tests have been performed in order to check our numerical procedure. First, we checked
the gauge invariance of the calculated results. Apart from the Feynman gauge which was normally
used, we repeated our calculations with the electron-electron interaction in the Coulomb gauge [see
Eq.~(\ref{cou})] and found perfect agreement. We also checked that varying the gauge of the
absorbed photon [parameter $G$ in Eq.~(\ref{1aa})] has no effect on the numerical results. Second,
we recalculated the contribution of the right diagram on Fig.~\ref{fig:dpi} by using a completely
different numerical technique, based on the explicit summation over the Dirac spectrum represented
by the finite basis set of $B$-splines \cite{shabaev:04:DKB}. Finally, we checked the
nonrelativistic limit of our calculations for the point-nucleus Coulomb potential against the
independent perturbation-theory calculations of Mikhailov {\em et al.} \cite{mikhailov:04} and
Amusia {\em et al.} \cite{amusia:75} and found good agreement.

\section{Results and discussion}
\label{sec:res}

In the present work, we study the process of the double $K$-shell photoionization for several
neutral medium-$Z$ atoms: magnesium ($Z = 12$), calcium ($Z = 20$), copper ($Z = 29$), and silver
($Z = 47$). The results of the calculation are presented in terms of the ratio of the
double-to-single $K$-shell ionization cross section,
\begin{align} \label{eq999}
{\cal R} = Z^2\, \frac{\sigma^{++}}{\sigma^+}\,,
\end{align}
as a function of the ratio of the incoming photon energy and the double $K$-shell photoionization
energy, $\omega/\omega_{\rm cr}$. The prefactor of $Z^2$ in the definition (\ref{eq999}) ensures
that the function ${\cal R}(\omega/\omega_{\rm cr})$ depend only weakly on the nuclear charge and
the degree of ionization of atom (ion); this fact is often termed as the universal scaling law of
double photoionization \cite{kornberg:94,mikhailov:09}. In our previous calculation for He-like
ions \cite{yerokhin:11:dpi}, we demonstrated that this scaling holds exactly in the {\em
nonrelativistic limit} and to the {\em first order} in perturbation theory, but is violated by the
relativistic effects.

In order to obtain numerical results for ${\cal R}(\omega/\omega_{\rm cr})$, we need first to
calculate the double $K$-shell ionization energy $\omega_{\rm cr}$ and the cross section of the
usual (single) $K$-shell photoionization $\sigma^+$. These calculations are relatively simple and
can be performed by many different methods. In the present investigation, however, we require that
$\omega_{\rm cr}$ and $\sigma^+$ were calculated within exactly the same approach as the double
photoionization cross section $\sigma^{++}$.

In Tables~\ref{tab:energy} and \ref{tab:sigma}, we present numerical results for the $K$-shell
double ionization energy $\omega_{\rm cr}$ and the cross section of the single $K$-shell
photoionization $\sigma^+$ for $\omega = 2\,\omega_{\rm cr}$, respectively. The columns labeled by
``CH$_1$'', ``CH$_2$'', and ``Coul'' display the data obtained in the present work with the
corresponding potential in the Dirac equation ($V_{\rm CH,1}$, $V_{\rm CH,2}$, and the Coulomb
potential, respectively). The Coulomb-potential values correspond to the case of He-like ion in the
independent particle model. The data in the ``MCDF'' column are obtained by the
Multiconfigurational Dirac-Fock method by using the RATIP package \cite{ratip}, whereas the data in
the last column are taken from the literature \cite{scofield}.

The single-photoionization cross sections ``CH$_1$'', ``CH$_2$'', and ``Coul'' are obtained within
the single-electron approximation, where the electron-electron interaction is accounted for
approximately either through the screening potential (for the CH$_1$ and CH$_2$ values) or is
totally ignored (for the Coulomb case). It is thus natural that these results are less accurate
than the ones obtained by the MCDF method. As might have been anticipated, the CH$_1$ potential is
the best choice among the three potentials considered. It is important that the difference between
the CH$_1$ and CH$_2$ values yields a reliable estimate of the residual electron-correlation
effects neglected by the effective single electron approximation.

\begin{table}
\caption{$K$-shell double ionization energy $\omega_{\rm cr}$, in keV.}
\begin{ruledtabular}
\label{tab:energy}
  \begin{tabular}{cccccc}
  $Z$ & CH$_1$ &  CH$_2$ & Coul  & MCDF  & Ref.~\cite{scofield}\\
    \hline\\[-5pt]
12  &          2.65   &  3.05	  &   3.93  & 2.79  & 2.61 \\
20 	&          8.12   &  8.79	  &   10.9  & 8.39  & 8.09 \\
29  &          18.0   &  19.0     &   23.2  & 18.4  & 18.0 \\
47  &          51.1   &  52.8	  &   62.0  & 51.9  & 51.3\\
  \end{tabular}
\end{ruledtabular}
\caption{Cross section of single $K$-shell photoionization $\sigma^+$ for the photon
energy $\omega = 2\,\omega_{\rm cr}$, in kbarn.}
\begin{ruledtabular}
\label{tab:sigma}
  \begin{tabular}{cccccc}
  $Z$ & CH$_1$ &  CH$_2$ & Coul & MCDF  & Ref.~\cite{scofield}\\
    \hline\\[-5pt]
12 	& 4.88	& 	  3.55	&	1.70  & 4.95 & 5.40\\
20  & 1.38	&	  1.15	&   0.61  & 1.40 & 1.52\\
29  & 0.575	&     0.510 &   0.288 & 0.635 & 0.649\\
47	& 0.183	&	  0.171 &	0.108 & 0.188 & 0.189\\
  \end{tabular}
\end{ruledtabular}
\caption{Ratio of the double-to-single $K$-shell photoionization cross sections
${\cal R} = Z^2\, \sigma^{++}/\sigma^+$, for the photon energy $\omega = 2\,\omega_{\rm cr}$.}
\begin{ruledtabular}
\label{tab:R}
  \begin{tabular}{ccccc}
  $Z$ & CH$_1$ &  CH$_2$ & Coul  & Coul (NR)\\
    \hline\\[-5pt]
12 	& 0.340	& 	  0.286	&	0.200 & 0.191 \\
20  & 0.316	&	  0.287	&   0.214 & 0.191 \\
29  & 0.323	&     0.305 &   0.239 & 0.191 \\
47	& 0.383	&	  0.373 &	0.318 & 0.191 \\
  \end{tabular}
\end{ruledtabular}
\end{table}

Let us now turn to the main objective of the present work, the ratio of the double-to-single
$K$-shell photoionization cross sections, ${\cal R}$. Our numerical results for ${\cal R}$ are
presented in Table~\ref{tab:R} for the photon energy $\omega = 2\,\omega_{\rm cr}$ and for the
three different potentials, CH$_1$, CH$_2$, and Coul. In the last column, we list also the
nonrelativistic limit of the Coulomb results. The striking feature of the presented comparison is
that the ratio ${\cal R}$ is much less sensitive to the choice of the potential than the single
photoionization cross section $\sigma^+$. This explains why we took great care in order to
calculate $\sigma^{++}$ and $\sigma^+$ fully consistently, i.e., both to the leading order in
perturbation theory with exactly the same screening potential in the Dirac equation (\ref{eq2}).
Importance of such consistency was implicitly acknowledged already in previous perturbation-theory
calculations \cite{amusia:75,mikhailov:04,yerokhin:11:dpi}, where results were presented solely in
terms of $\cal R$, and not in terms of $\sigma^{++}$.

From the comparison in Tables~\ref{tab:energy}--\ref{tab:R}, we see that the difference between the
CH$_1$ and CH$_2$ results (which we use for estimating the magnitude of the residual electron
correlation effects) monotonically decreases with the increase of the nuclear charge. It is
consistent with what one might have anticipated: the single-electron approximation usually performs
the better, the heavier the atom under consideration.

The nonrelativistic Coulomb values of ${\cal R}$ in Table~\ref{tab:R} are exactly the same for all
atoms. This demonstrates the universal scaling law of the {\it nonrelativistic} double
photoionization \cite{kornberg:94,mikhailov:09}. As seen from the table, the nonrelativistic
scaling is violated by the relativistic effects, so that the relativistic Coulomb results depend on
nuclear charge.

\begin{figure*}[tb]
  \vspace*{5.0cm}
  \centerline{\includegraphics[width=0.8\textwidth]{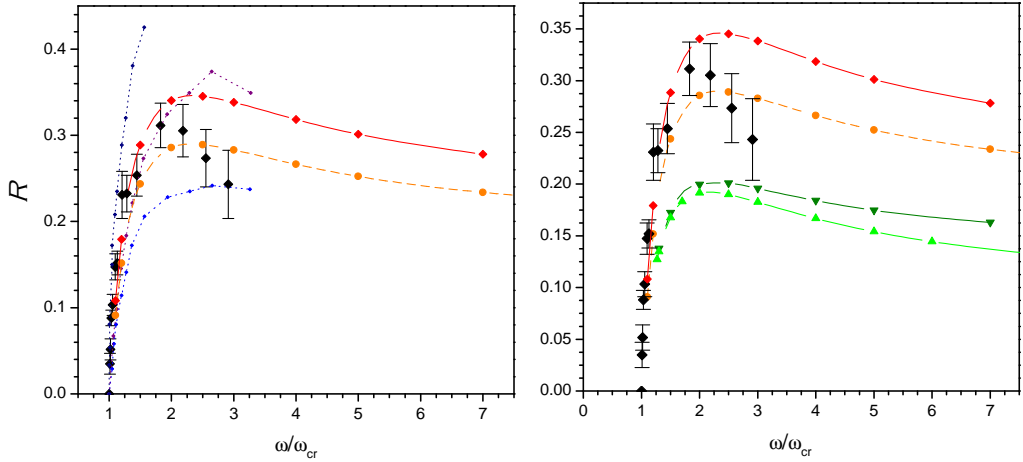}}
\caption{(Color online) Double $K$-shell photoionization cross section for neutral magnesium (Mg, $Z = 12$).
The double-to-single $K$-shell ionization ratio (${\cal R} = Z^2 \sigma^{++}/\sigma^+$) is plotted
as function of the incoming photon energy $\omega$ in units of
the double $K$-shell photoionization energy $\omega_{\rm cr}$. The results of the present
calculation are plotted by solid line (red) for the CH$_1$ potential and by dashed line (orange)
for the CH$_2$ potential, respectively. The experimental results of
Refs.~\cite{hoszowska:09,hoszowska:10} are shown by black diamond points.
On the left panel, the calculations by Kheifets {\em et al.} \cite{kheifets:09} in
length, velocity, and acceleration gauge are shown by dotted lines (navy, purple, and blue, respectively).
On the right panel, the present theory and experiment \cite{hoszowska:09,hoszowska:10} for neutral
magnesium are compared with the computations for He-like magnesium ion. The relativistic
results for the He-like ion are shown by down triangle points (dark green line); the nonrelativistic
calculation is shown by up triangle points (light green line).
\label{fig:Mg}}
\end{figure*}

We now discuss our results for selected atoms in more detail. Fig.~\ref{fig:Mg} presents our
calculation for magnesium. The left panel compares our results with the nonrelativistic
calculations in the length, velocity, and acceleration gauge by Kheifets {\em et al.}
\cite{kheifets:09} as well as the experiments by Hoszkowska {\em et al.}
\cite{hoszowska:09,hoszowska:10}. Our results are shown for the CH$_1$ (red solid line) and CH$_2$
(orange dashed line) potentials. In addition, the right panel of this figure displays our
calculations for the corresponding He-like ion (Coulomb potential), presenting the relativistic
results (dark green, down triangle points) and nonrelativistic results (light green, upper triangle
points). The results for the He-like ions are equivalent to ones obtained in our previous work
\cite{yerokhin:11:dpi}.

From the comparison in the right panel of Fig.~\ref{fig:Mg}, we can identify the magnitude of
various effects. The difference between the nonrelativistic and the relativistic curves for He-like
ion shows the effect of relativity, whereas the difference between the CH$_2$ results for the
neutral-atom and the relativistic results for He-like ion identifies the effect of the outer
electrons. The difference between the CH$_1$ and CH$_2$ curves might be interpreted as the effect
of higher-order interactions between the two $K$-shell electrons. We may therefore conclude that
for magnesium, the relativistic effects are not very prominent but the interaction with the outer
shells increases the ratio ${\cal R}$ by about 50\%{} and thus cannot be neglected.

The results of our calculation for magnesium agree reasonably well with the experimental data
\cite{hoszowska:09,hoszowska:10} for energies up to the maximum of the curve ($\omega/\omega_{\rm
cr} \approx 2$) but deviate noticeably for higher photon energies. Good agreement is also observed
with the nonrelativistic results by Kheifets and coworkers~\cite{kheifets:09} obtained in the
velocity gauge, while their results in acceleration and length gauge deviate significantly both
from our predictions as well as from experimental data.

\begin{figure*}[tb]
  \vspace*{5.0cm}
  \centerline{\includegraphics[width=0.8\textwidth]{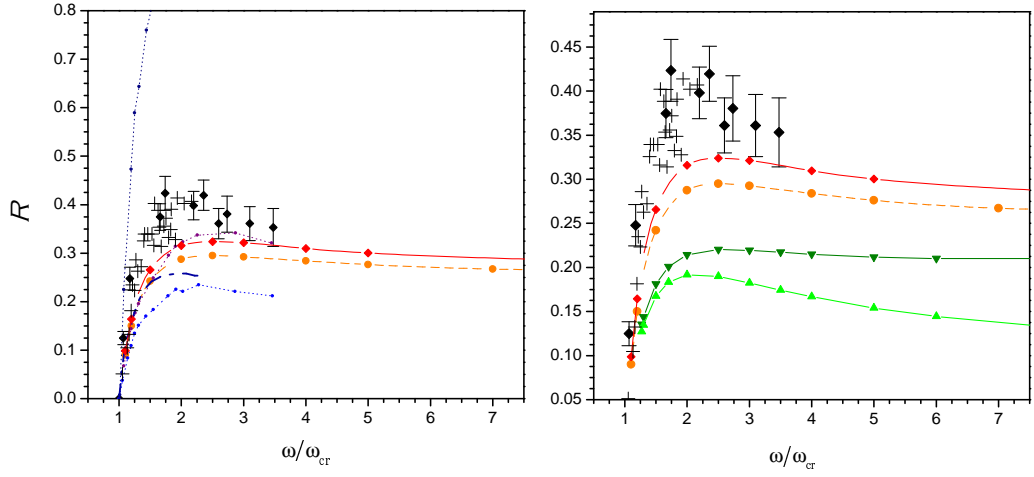}}
\caption{(Color online) The same as Fig.~\ref{fig:Mg} but for neutral calcium (Ca, $Z = 20$).
The crosses denote the experimental results of Oura {\em et al.} \cite{oura:02}.
On the left panel, the violet dash-dotted line denotes the calculation by Mikhailov {\em et al.}
\cite{mikhailov:04}.
  \label{fig:Ca} }
\end{figure*}

Fig.~\ref{fig:Ca} displays our results for calcium and compares them with previous calculations
\cite{kheifets:09,mikhailov:04} and experiments \cite{hoszowska:09,hoszowska:10,oura:02}. We
observe that for this element, neither of calculations agrees well with experiment. The
computations by Kheifets {\em et al.} \cite{kheifets:09} exhibit a very strong gauge dependence. At
the same time, similarly as for magnesium, their results in velocity gauge are found to be in good
agreement with our values. The computations by Mikhailov {\em et al.} \cite{mikhailov:04} agree
with our results and with Kheifelts's velocity-gauge data for small photon energies of
$\omega/\omega_{\rm cr} \lesssim 1.5$ but predict significantly smaller values of ${\cal R}$ at
higher photon energies.

\begin{figure*}[tb]
  \vspace*{5.0cm}
  \centerline{\includegraphics[width=0.45\textwidth]{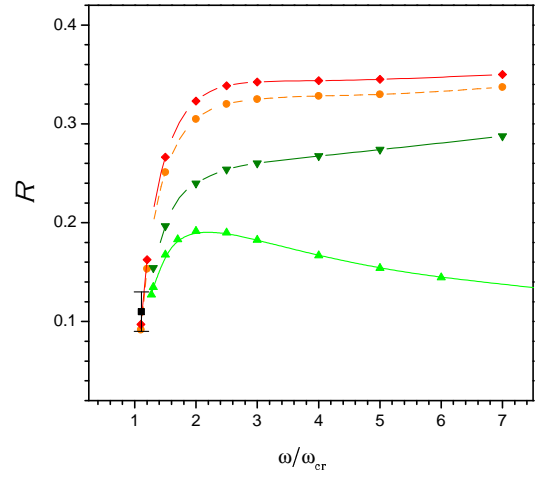}}
\caption{(Color online) The same as Fig.~\ref{fig:Mg} but for neutral copper (Cu, $Z = 29$).
The experimental result is by Diamant {\it et al.}~\cite{diamant:00:Cu}.  \label{fig:Cu}
}
\end{figure*}

Fig.~\ref{fig:Cu} presents the ratio $\cal R$ for copper. Results of our relativistic calculation
for the neutral atom with the screening potentials CH$_1$ and CH$_2$ (the red solid line and the
orange dashed line, respectively) are compared with our relativistic calculation for the He-like
ion (dark green line, down triangle points) and the corresponding nonrelativistic calculation
(light green line, upper triangle points). Here, our results agree well with the only available
experimental point \cite{diamant:00:Cu} in the near-to-threshold region.

\begin{figure*}[tb]
  \vspace*{5.0cm}
  \centerline{\includegraphics[width=0.45\textwidth]{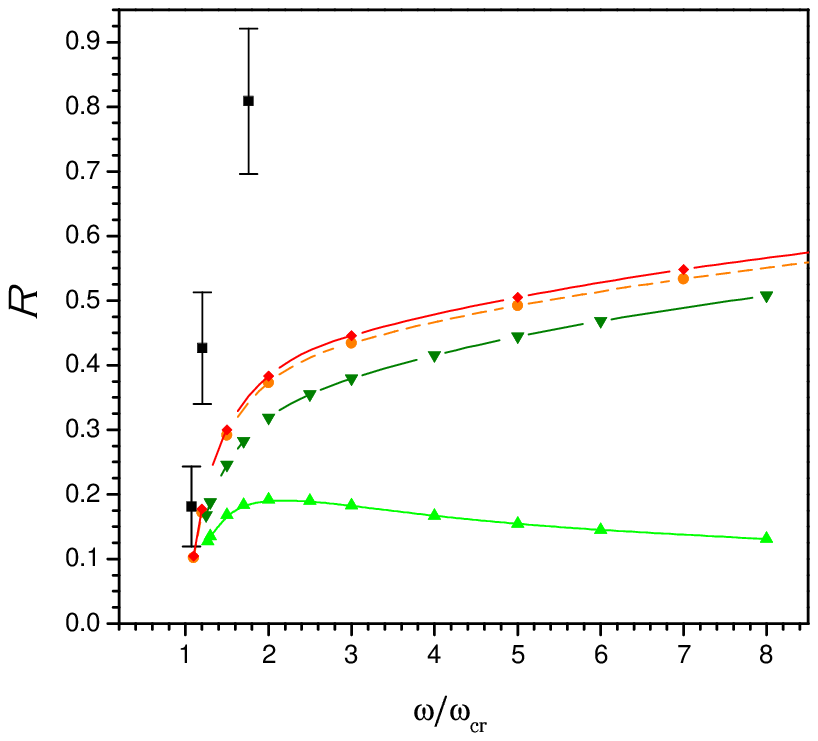}}
\caption{(Color online) The same as in Fig.~\ref{fig:Cu}, but for neutral silver (Ag, $Z = 47$).
The experimental points are by Kantler {\em et al.} \cite{kantler:06}.
  \label{fig:Ag} }
\end{figure*}

Finally, Fig.~\ref{fig:Ag} presents theoretical and experimental results for silver, which is the
heaviest atom for which measurements of the double $K$-shell ionization have been performed. We
observe that two of the three experimental points reported in Ref.~\cite{kantler:06} disagree
strongly with the theory. Comparison of theoretical curves for neutral atom and He-like ion shows
that the influence of the outer-shell electrons is rather weak in this case, so it seems unlikely
that the residual electron correlation can explain the discrepancy. The relativistic effects are
large and change the form of the curve remarkably \cite{yerokhin:11:dpi}, but they are not large
enough in order to bring theory into agreement with experiment.

More generally, our calculations show that the relative influence of the outer-shell electrons on
${\cal R}$, being significant for low-$Z$ systems, gradually decreases as $Z$ increases. This is
what one might have anticipated, having in mind that the characteristic photon energy scales as
$Z^2$, so that the interaction of electrons with the photon should become increasingly localized
around the nucleus when $Z$ increases.

It is interesting that for small and medium photon energies, $\omega/\omega_{\rm cr} \lesssim 2$,
the relativistic results for ${\cal R}$ exhibit essentially the same scaling law for neutral atoms
as the one reported previously for the nonrelativistic He-like ions
\cite{kornberg:94,mikhailov:09}. In particular, the relativistic theoretical values of ${\cal R}$
for $\omega/\omega_{\rm cr} = 2$ are nearly the same for all neutral atoms considered in the
present work (about 0.35, see Table~\ref{tab:R}). This is because, for low-$Z$ atoms, the
electron-correlation effects are large and relativistic effects are small, whereas for medium-$Z$
atoms, it is vice versa. Since both effects increase the ratio ${\cal R}$, the sum of them leads to
a nearly uniform (in $Z$) enhancement of the familiar nonrelativistic He-like curve.

The situation becomes drastically different for large photon energies, $\omega/\omega_{\rm cr}
\gtrsim 2$. In this region, the ${\cal R}$ curve for light atoms decreases gradually, similarly to
what is found for the well-studied nonrelativistic helium case. For heavy atoms, however, the
relativistic effects change this behaviour. Already for copper the familiar nonrelativistic peak of
the curve around $\omega/\omega_{\rm cr} = 2$ disappears (Fig.~\ref{fig:Cu}), while for silver the
curve becomes monotonically increasing (Fig.~\ref{fig:Ag}). Our previous calculation
\cite{yerokhin:11:dpi} suggests that the relativistic curve increases monotonically also at higher
energies, thus changing the asymptotic behaviour with regard to the nonrelativistic theory.

We mention that in all experimental studies \cite{hoszowska:09,hoszowska:10,kantler:06}, the
experimental data for ${\cal R}$ were fitted to semi-empirical curves assuming the nonrelativistic
behaviour in the high-energy limit, which is not fully justified in the case of medium-$Z$ atoms,
according to our calculations.

\section{Conclusion}

We performed calculations of the double $K$-shell photoionization cross section for several
neutral, medium-$Z$ atoms from magnesium ($Z = 10$) up to silver ($Z = 47$). Our fully relativistic
approach accounts for all multipoles of the absorbed photon as well as the retardation (the
frequency dependence) of the electron-electron interaction. The electron-electron interaction was
taken into account rigorously to the leading order of perturbation theory. The higher-order
electron-electron interactions (in particular, with the outer-shell electrons) were treated
approximately by means of some screening potential in the Dirac equation. The approach of this work
is strictly gauge invariant, and this was utilized in order to cross check the computational
procedure.

The results of our computations are in reasonable agreement with experimental data
\cite{hoszowska:09,hoszowska:10,diamant:00:Cu} for the light elements, but they disagree strongly
with the measurement of Ref.~\cite{kantler:06} for silver. The reason of this disagreement is
unknown.

Our calculations predict large relativistic effects for copper ($Z = 29$) and for heavier atoms as
well as at large photon energies (more than twice larger than the double $K$-shell ionization
energy). For these energies, the shape of the ${\cal R}$ curve is changed quite remarkably due to
relativity. This prediction cannot be presently tested against experiment, as there have not been
any direct measurements of double photoionization in this region so far.

%
%

%
%

%
%

%
%

\section*{Acknowledgement}

The work reported in this paper was supported by BMBF under Contract No.~05K13VHA.


\end{document}